# Can One Design a Series of Brains for Neuromorphic Computing to solve complex inverse problems?

**Mingyong Zhou**

School of Computer Science and Communication Engineering

GXUST
Liuzhou, China
Zed6641@hotmail.com

**Abstract.**

In this position paper, we present a discussion on neuromorphic computing and especially the learning/training algorithm to design a series of brains with different memristive values to solve complex ill-posed inverse problems based on a Finite Element(FE) method. First, the neuromorphic computing is addressed and we focus on a type of memristive circuit computing that falls into the scope of neuromorphic computing. Secondly based on reference [1] in which the complex dynamics of the complex memristive circuit was studied, we design a method and an approach to train the memristive circuit so that the memristive values are optimally obtained. Instead of using traditional back propagation (BP) algorithm that could accumulate the computing errors during the propagation, we derive a set of linear equations to obtain the optimal memristive values based on the Finite Element analysis of the original complex dynamical differential equations. We outline this method in this section. Last but not least, we show how this neuromorphic computing learning/training algorithm can be applied in two inverse problem applications, that is, Electrical Impedance Tomography(EIT) and Blind Speech Separations.

By converting the dynamical differential equations that describe the memristive circuit into linear equations by Finite Elements(FE) method, we are able to focus on the discussions on the learning/training algorithms based on the linear algebra method. This method proposed in this paper is generic and constructive as long as the training samples data are large enough and the differential equations as well as boundary conditions that describe the neuron dynamics are certain.

**Keyword**: Neuromorphic computing, Memristive circuit, Electrical Impedance Tomography (EIT) , Finite Element(FE),

# 1 Introduction

Neuromorphic computing is a new direction in computing research that addresses to solve complex computing problems that are difficult to be solved by traditional methods. By carefully selecting the structures of neuron network types such as RNN or CNN, neuron networks are trained first by using large samples to obtain the optimal synapses that are either constant or time varying for a special application scenario. Then the neuron network can be applied into new tasks in a special application such as a particular inverse problem of Electrical Impedance Tomography or Blind Separations etc. Considering the complexity of various inverse problems that are either linear or highly non-linear , neuromorphic computing provides a new direction and an approach that could tackle a complex inverse problem that has been challenging for researchers for years.

In this paper based on reference[1] we present and investigate a type of memristive circuit that falls into the scope of neuromorphic computing . The dynamics of the complex network is investigated in [1]. Based on their work on differential equations describing the dynamics, we derive a set of linear equations based on Finite Element method , thus derive a new method of designing the training algorithms to optimally obtain the memristive values that could be constant or time varying. We outline this method and show how this proposed approach works in this position paper. We as well highlight and compare with back propagation (BP) method that are widely used as training algorithms. As the applications of the approach and method, we particularly demonstrate how to solve the EIT and Blind separation inverse problems.

We mainly devote our discussions in this paper to the training algorithms design based on FE method and show how it will work in a few of particular inverse problem applications.

# 2 Differential Equation for Neuromorphic Computing Dynamics

We try to establish a dynamical differential equations in time domain regarding the menristive values, as well as spatial differential equations describing the spatial distributions of the menristive values. We will start from the differential equations of the memristive circuit's dynamics based F. Caravelli et al's work in reference [1]. However the modelling of the whole memristive dynamics differential equations both in spatial and time domain will be the key work for us in the first place.

F. Caravelli et al's work in reference [1] involves the following type of dynamical differential equation:

$$\frac{d\vec{W}}{dt} = \alpha \vec{W} - \frac{1}{\beta} f(W),$$

$$f(W) = JA^t (\overline{A}A^t + (r-1)\overline{A}WA^t)^{-1} A\vec{S}(t)$$

(2.1)

Where $\vec{W}$ denotes a vector whose memristive values are time-varying in nature inside the memristive circuit, $\vec{S}(t)$ denotes the vector elements formed from the voltages imposed on the edges of the memristive circuit boundary, all other parameters notations are the exact the same as in reference [1]. In addition to (2.1), in fact a set of spatial differential equations based on Maxwell theory can be derived as well, which is closely related to Electrical Impedance Tomography(EIT). Instead of presenting in this section, we describe them in a separate **Section 5** later.

We wish to highlight that the differential equation represented by equation (2.1) is actually the type of Cohen-Grossberg proposed in 1983, the dynamical stability of which is well described in reference [5].

In fact we can alternatively switch between spatial differential equations in **section 4&5** and time dynamical differential equation in (2.1) to obtain the memristive distributions for a given time sequence instead of only solving the Maxwell partial differential equations at all given time sequences that will cost huge computation resources. If both the dynamical equation (2.1) and spatial differential equations in section 5 accurately describe the menristive circuit and they are consistent with each other in describing the time and spatial dynamics of the menristive circuit, then it is always possible to find a solution based on (2.1) and the Maxwell equations in **section 4&5**.

## 3 Finite Element(FE) method to solve differential equations

Given a set of partial differential equations with boundary conditions, it is well known that Finite Method can be applied to solve the differential equations. This generic method is well established in mechanics analysis and is briefly outlined in this section. We however focus on the application of the FE method into our training algorithm design in the following section.

For a 3D bounded by Ω, we can divide the 3D bounded space Ω with the following triangle with K,M,N,L and coordinates represented by ($x_k,y_k,z_k$), ($x_M,y_M,z_M$), ($x_N,y_N,z_N$), ($x_L,y_L,z_L$).

**Figure 3.1: Finite Element Triangle**

The derivation of the Finite Element(FE) method is too tedious but is a well-known and well used method in mechanics engineering. By diving the bounded space Ω with small triangles and enough numbers of triangles, for the ith injection of currents we can formulate a linear equations as follows:

$$S_i \Phi_i = F_i \quad (3.1)$$

where $F_i$ is a vector that is unrelated to memristive values $\rho$ and is related only to input signals, $\Phi_i$ is a vector formulated by both observable output signals and unobserved signals at each finite elements inside and $S_i$ is a matrix formulated by the memristive values $\rho$ of each triangle elements. One should note that equation (3.1) is derived from differential equations in previous section, thus the vector values and matrix values in (3.1) is highly dependant upon the differential equations in section 2. Now the inverse problem becomes to estimate the values of Matrix $S_i$, given the partially measured and observed vector $\Phi_i$ and vector $F_i$ that is formulated as the input signals. Usually this is an ill-posed inverse problem and regular operators must be used to obtain a solution for matrix $S_i$. In the following section, however we will show that this ill-posed issue will not bother us to make use of FE method and its linear equations to design a training algorithm so that we can propose a series of neuron networks or "brains" that are composed of totally different time-varying memristive values set but can solve the same type of particular inverse problem!

## 4 Learning/Training method to solve the complex inverse problem

Starting from equation (3.1), we now are able to have a discussion on how to design a training procedure: if we interpret index i as the discrete time sequence, at each ith time we obtain a set of linear equations denoted by (3.1), where Fi denotes a vector formulated by input signals and Si denotes a vector whose partial values are formulated by observable output signals. Now we can train the neuron network by a series of input signals that lead to a series vectors $F_1, F_2, ..., F_N$ as known vectors. Then we can observe the output series that however formulate only parts of values in vectors $\Phi_1, \Phi_2...\Phi_N$. Matrix $S$ contains at ith time all memristive values that needs to be obtained optimally. If we denote the various input signals as F1, F2...and output observable signals as $\Phi_1, \Phi_2...\Phi_N$, at the ith time a set of linear equations can be catenated as follows

$$S[\Phi_1, \Phi_2...\Phi_N] = [F_1, F_2,...,F_N] \qquad (4\text{-}1)$$

One should note that equation (4.1) exists for any given time. Note that due to the Finite Element method, observable output signals are included in $\Phi_1, \Phi_2...\Phi_N$ but some values defined by Finite Element method inside $\Phi_1, \Phi_2...\Phi_N$ are still missing which means that they are not observable by output signals. However since we are only interested in training a neuron network and obtain the memristive values as one of the many solutions, this does not bother us to proceed. A straightforward method is to set pre-defined values to those unobserved values in $\Phi_1, \Phi_2...\Phi_N$ so that the matrix formed by $[\Phi_1, \Phi_2...\Phi_N]$ is non-singular, in this way we can optimally estimate the matrix S that contains all the memristive values at ith time. At each time sequence, we repeat this process so that at each time sequence a time varying set of memristive values are optimally obtained. In other words the training process is completed this way to obtain the memristive values at different time sequence. By equation (4.1) we in fact assume that the memristive values only change with time, they do not change with different input signals, which is an only one assumption made in our algorithm design.

We note by invoking Finite Element(FE) method, in fact we convert a nonlinear problem into a linear problem while the time-varying problem is actually converted into a series of time-invariant linear equations at different given time sequence. In this way, we are actually proposing a series of "brains" each of which can equally, at least in theory, tackle a particular inverse problem in a particular application. One obvious advantage, as compared to the well-used Back Propagation(BP) method is that our algorithm is based on a full analysis of linear equations instead of the iterative procedure that could help accumulate the errors during the propagation. Another important feature of our algorithm method is that as long as the linear numerical stability is guaranteed , which is defined by the linear equation in (4.1) there are no such problems as gradient explosion or gradient vanish that happened in other gradient learning/training algorithms. Our learning /training algorithm design is thus straightforward , linear , transparent and analytical.

## 5 Electrical Impedance Tomography(EIT) and other inverse problems

In reference [2], EIT problem is defined and described. According to the Maxwell theory and Ohm law, we can get the description like this:

$$-\rho^{-1}\nabla\phi = \vec{J} \quad (5.1)$$

In this equation, ρ represents the distribution function of impedance, ϕ means potential distribution, J is the function of boundary electric current density.

Due to there is no electric current through interior of biological tissue, so we get this:

$$\nabla \cdot \vec{J} = 0 \quad (5.2)$$

Combined (1-1) and (1-2), we can get equation :

$$\nabla \cdot \rho^{-1}\nabla\phi = 0 \quad (5.3)$$

This partial differential equation should satisfy the Dirichlet boundary condition

$$\phi\vert_{\partial\Omega} = V \quad (5.4)$$

In that, $\Omega$ is the area where the object in. One is required to estimate all the impedance values inside $\Omega$ which is actually a complex inverse problem considering the non-linearity and differential equations. In reference [3] a factorization method is proposed based on an investigation of well-known MUSIC algorithm in signal processing.

Now with the introduction of section 4 , we can apply a neuromorphic computing method to train the memristive values using large samples with known impedance configurations . Thus we are provided with a series of "brains" after the training procedure using large samples and we can then use such trained "brains" to solve new tasks in electrical impedance tomography problem. One should note that our algorithm is generic and constructive. In another less complex inverse problem, one is required to separate multiple speech signals from a mixed one under a noisy scenario. One can similarly design a training algorithm to solve such blind separation problem.

## 6 Further discussions

In this paper based on reference[1] ,we present a learning/training algorithm to design a series of "brains" with different set of memristive values to solve a particular type of inverse problem based on Finite Element(FE) method. First the differential equations describing the complex dynamics are presented , secondly a set of linear equations are derived based on Finite Element method , and we show how to make use of the set of linear equations to train the brains so that optimal memristive values are obtained at different time sequence. Lastly we outline the assumptions we made on our algorithm and show the advantage compared with BP algorithm. An example of application in electrical impedance tomography is described on how to apply our proposed training algorithm.

However our algorithm is based on Finite Element (FE) method, therefore the accurate mathematical differential equations as well as the boundary conditions must be well described about the neuron networks so that FE linear equations can be derived correctly and exactly. If the differential equations describing the neuron network are not accurate and complete, the trained " brains" based on FE method may not work well for practical problems. For more complicated neuron networks than memristive circuit such as CNN, RNN or their revised types, accurate dynamical differential modeling as well as investigation on their dynamical stability and their Lyapunov function is necessary. This is an important but inevitable task of mathematical modelling which is not covered in this paper because this topics is related to many deep neuron network and their revised types. In this paper however we propose a clue and a method to design the "brains" with memristive circuit based on the condition that the dynamical properties and stability are well described by differential equations and training samples are large enough.

At the end of preparing this paper, we noted a discussion in reference[4] in which the difference between neuromorphic computing and deep neuron network is discussed. As indicated in [4], the structural and operational differences between the two are not fundamental, which could be a milestone to indicate that the investigations on neuronmorphic computing actually merge into the deep learning neuron networks framework. In other words, the research results from neuromorphic computing , including the method presented in this paper, can be mapped into deep learning neuron networks framework and vice versa.

# 7 Acknowledgement